\documentclass[doublecol]{epl2}
\usepackage{bm}% bold math
\usepackage{amsmath}
\usepackage{amssymb}

\begin{document}

\title{Autonomous elastic microswimmer}
\shorttitle{Autonomous three-sphere microswimmer}
%Insert here a short version of the title if it exceeds 70 characters

\author{Katsutomo Era\inst{1}, Yuki Koyano\inst{2}, 
Yuto Hosaka\inst{1}, Kento Yasuda\inst{1},
Hiroyuki Kitahata\inst{3} \and Shigeyuki Komura \inst{1}}
\shortauthor{K. Era \textit{et al.}}

\institute{
\inst{1} Department of Chemistry, Graduate School of Science,
Tokyo Metropolitan University, Tokyo 192-0397, Japan \\
\inst{2} Department of Physics, Graduate School of Science, Tohoku University, Sendai 980-8578, Japan \\
\inst{3} Department of Physics, Graduate School of Science, Chiba University, Chiba 263-8522, Japan
}


\abstract{
A model of an autonomous three-sphere microswimmer is proposed by implementing a coupling effect 
between the two natural lengths of an elastic microswimmer. 
Such a coupling mechanism is motivated by the previous models for synchronization phenomena in 
coupled oscillator systems.
We numerically show that a microswimmer can acquire a nonzero steady state velocity and 
a finite phase difference between the oscillations in the natural lengths.
These velocity and phase difference are almost independent of the initial phase difference.
There is a finite range of the coupling parameter for which a microswimmer can have an autonomous 
directed motion.
The stability of the phase difference is investigated both numerically and analytically in order to 
determine its bifurcation structure.
}

\maketitle
%\baselineskip=18pt

%%%%%%%%%%%%
\section{Introduction}
%%%%%%%%%%%%

Microswimmers are small machines that swim in a fluid and they are expected to be used in microfluidics 
and microsystems~\cite{Lauga09}.
Over the length scale of microswimmers, the fluid forces acting on them are dominated by the frictional 
viscous forces.
By transforming chemical energy into mechanical energy, however, microswimmers change their 
shape and move efficiently in viscous environments.
According to Purcell's scallop theorem, reciprocal body motion cannot be used for locomotion in a 
Newtonian fluid~\cite{Purcell77,Lauga11}.
As one of the simplest models exhibiting nonreciprocal body motion, Najafi and Golestanian proposed 
a three-sphere swimmer (NG swimmer)~\cite{Golestanian04,Golestanian08}, in which three in-line spheres 
are linked by two arms of varying length.
In recent years, such a swimmer has been experimentally realized by using colloidal beads manipulated 
by optical tweezers~\cite{Leoni09}, or ferromagnetic particles at an air-water 
interface~\cite{Grosjean16,Grosjean18}.

Recently, some of the present authors have proposed a generalized three-sphere microswimmer model 
in which the spheres are connected by two harmonic springs, i.e., an elastic 
microswimmer~\cite{Yasuda17}.  
Compared with the NG swimmer, the main difference is that the natural length of each spring 
(rather than the arm length) is assumed to undergo a prescribed cyclic motion. 
A similar model was proposed by other people~\cite{Dunkel09,Pande15,Pande17}.
We have analytically obtained the average swimming velocity as a function of the frequency of 
cyclic change in the natural length~\cite{Yasuda17}.
Using this model, we have also discussed the hydrodynamic interaction between two elastic 
swimmers~\cite{Kuroda19} and a thermally driven elastic microswimmer~\cite{Hosaka17,Sou19,Sou20}.

In the above three-sphere microswimmer models, either the arm lengths (NG swimmer) or the natural 
lengths of the springs (elastic swimmer) are assumed to undergo prescribed cyclic motion.
Such active motions can lead to a net locomotion if the swimming strokes are nonreciprocal. 
In these models, the average swimming velocity fluid is purely 
determined by the frequency and the phase difference of the prescribed motions~\cite{Golestanian04,Golestanian08,Yasuda17,Kuroda19}.

On the other hand, it is beneficial for a microswimmer if the swimming velocity is autonomously 
determined by itself rather than being imposed externally.
Moreover, a sophisticated microswimmer requires a feedback control system in order 
to regulate the switching between the static and swimming states by tuning the system parameters.
For a macroscopic quadruped robot (not a swimmer), it was demonstrated that the communication between legs during 
movements is essential for interlimb coordination in quadruped walking~\cite{Owaki12,Fukuhara18}.
A similar mechanism is also useful for the locomotion of a microswimmer.

\begin{figure}[tbh]
\centering
\includegraphics[scale=0.3]{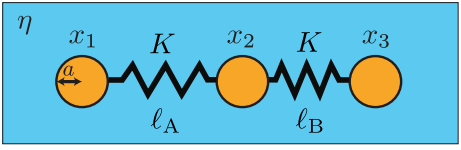}
\caption{
An autonomous elastic microswimmer in a viscous fluid characterized by the shear 
viscosity $\eta$. 
Three identical spheres of radius $a$ are connected by two harmonic springs characterized by 
the spring constant $K$.
The time-dependent positions of the spheres are denoted by $x_1$, $x_2$, and $x_3$ 
which evolve in time according to eq.~(\ref{ssokudo}).
The time-dependent natural lengths of the springs are denoted by $\ell_{\rm A}$ and $\ell_{\rm B}$ 
whose dynamics is described by eqs.~(\ref{ellA}) and (\ref{ellB}), respectively, 
whereas the corresponding phases $\theta_{\rm A}$ and $\theta_{\rm B}$ obey 
eqs.~(\ref{time_evolution_phase1}) and (\ref{time_evolution_phase2}), respectively.}
\label{model}
\end{figure}

\begin{figure}[tbh]
\centering
\includegraphics[scale=0.23]{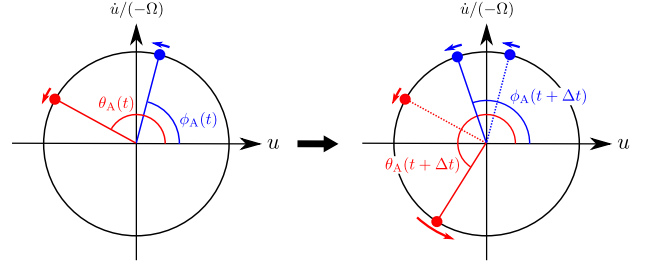}
\caption{
Dynamics of $\theta_{\rm A}$ describing the phase of the natural length [see eq.~(\ref{ellA})]
and $\phi_{\rm A}$ describing the mechanical phase [see eq.~(\ref{phiA})].
When $\theta_{\rm A} > \phi_{\rm A}$ at $t$, as shown in the left figure, and when $\alpha>0$ in 
eq.~(\ref{time_evolution_phase1}), the velocity $\dot{\theta}_{\rm A}$ becomes larger at a later time 
$t + \Delta t$, as shown in the right figure.
As a result, the difference between $\theta_{\rm A}$ and $\phi_{\rm A}$ also increases at $t + \Delta t$.
A similar dynamics occurs also for $\theta_{\rm B}$ and $\phi_{\rm B}$.}  
\label{phasedynamics}
\end{figure}

In this letter, extending the mechanism of an elastic swimmer~\cite{Yasuda17}, we propose a new type of 
three-sphere swimmer which can autonomously determine its velocity.
In order to implement such a control mechanism, 
we introduce a coupling between the two natural lengths of an elastic microswimmer by 
using the interaction adopted in the Kuramoto model for coupled oscillators~\cite{Kuramoto84,Pikovsky01,Strogatz03,Ritort05}.
Importantly, the proposed microswimmer acquires a steady state velocity and a finite phase difference in 
the long-time limit without any external control. 
The steady state velocity can be mainly tuned by changing the coupling parameter in the model.
Moreover, we investigate the condition that a microswimmer can attain an autonomous locomotion, 
and further perform a linear stability analysis of the steady state.

Synchronization phenomena are widely observed in active biological systems such as 
flagella and cilia~\cite{Golestanian11,Uchida17}.
In particular, synchronization of a pair of flagella in \textit{Chlamydomonas} was observed 
experimentally~\cite{Polin09,Goldstein09}.
For a three-sphere model of \textit{Chlamydomonas} in which the spheres representing the flagella 
move on circular trajectories relative to the body sphere~\cite{Friedrich12,Bennett13}, 
the two flagella can synchronize due to the local hydrodynamic friction forces~\cite{Friedrich12}.
On the other hand, hydrodynamic interaction between the flagella is indispensable for the net swimming.

%%%%%%%%%%%%%%%%%%%%%%%%%%
\section{Model of an autonomous microswimmer}
%%%%%%%%%%%%%%%%%%%%%%%%%%

As schematically shown in fig.~\ref{model}, the present model consists of three hard spheres 
of the same radius $a$ connected by two harmonic springs characterized by the spring constant $K$.
The total elastic energy is given by

\begin{align}
E = \frac{K}{2}(x_2 - x_1 - \ell_{\rm A})^2 + 
\frac{K}{2}(x_3 - x_2 - \ell_{\rm B})^2,
\end{align}
where $x_i(t)$ ($i=1, 2, 3$) are the positions of the three spheres in a one-dimensional 
coordinate system and we assume $x_1<x_2<x_3$ without loss of generality.
In the above, $\ell_{\rm A}(t)$ and $\ell_{\rm B}(t)$ are the natural lengths of the springs 
and their dynamics will be explained later [see eqs.~(\ref{time_evolution_phase1}) and 
(\ref{time_evolution_phase2})].
Each sphere exerts a force on the viscous fluid of shear viscosity $\eta$ and experiences 
an opposite force from it.

Denoting the velocity of each sphere by $\dot x_i=dx_i/dt$ and the force acting on each sphere by $f_i$, 
we can write the equations of motion of each sphere as~\cite{Yasuda17,Kuroda19}

\begin{equation}
\dot x_i  = \sum_{j=1}^{3}M_{ij}f_j,
\label{ssokudo}
\end{equation}
where the three forces $f_i$ are given by  

\begin{equation}
f_i=-\frac{\partial E}{\partial x_i}.
\label{forcevel}
\end{equation}
Here the details of the hydrodynamic interactions are taken into account through the mobility 
coefficients $M_{ij}$.
Within Oseen's approximation, which is justified when the spheres are considerably far from each other
($a \ll \vert x_i-x_j \vert$), the expressions for the mobility coefficients $M_{ij}$ can be written as

\begin{equation}
M_{ij}=
\begin{cases}
\dfrac{1}{6\pi\eta a} & i=j, 
\\[1.5ex]
\dfrac{1}{4\pi\eta \vert x_i-x_j \vert} & i\neq j.
\end{cases}
\label{joken}
\end{equation}
The force-free condition, $f_1+f_2+f_3=0$, is automatically satisfied in the present 
model~\cite{Yasuda17,Kuroda19}. 
We define the center-of-mass position of a microswimmer by $X(t)=(x_1+x_2+x_3)/3$ and the 
swimming velocity of the whole object by $V(t) = \dot{X}(t)$.

\begin{figure}[tbh]
\centering
\includegraphics[scale=0.23]{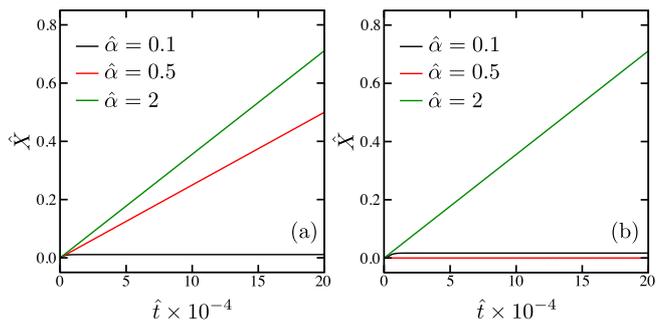}
\caption{
The plots of dimensionless center-of-mass position $\hat{X}$ of an autonomous three-sphere microswimmer 
as a function of dimensionless time $\hat{t}$ for $\hat{\Omega}=0.1$ when the initial phase differences are 
(a) $\delta_0=-\pi/2$ and (b) $\delta_0=-39\pi/40$.
In both plots, the dimensionless coupling parameter is chosen as $\hat{\alpha}=0.1$ (black),  
$0.5$  (red), and $2$ (green).
}
\label{fig:d-time}
\end{figure}

Next, we consider that the two natural lengths of the springs undergo the following cyclic 
changes in time~\cite{Yasuda17,Kuroda19}

\begin{align}
\ell_{\rm A}(t) & =\ell+d \cos \theta_{\rm A}(t),
\label{ellA} \\
\ell_{\rm B}(t) & =\ell+d \cos \theta_{\rm B}(t),
\label{ellB}
\end{align}
where $\ell$ is the constant natural length, $d$ is the oscillation amplitude, $\theta_{\rm A}(t)$ 
and $\theta_{\rm B}(t)$ are the time-dependent phases.
The most important aspect of our model is that $\theta_{\rm A}(t)$ and $\theta_{\rm B}(t)$
are affected by the relative positions and the velocities of the three spheres.
We employ the following time-evolution equations for $\theta_{\rm A}(t)$ and 
$\theta_{\rm B}(t)$ which are often used to describe synchronization phenomena~\cite{Kuramoto84}

\begin{align}
\dot{\theta}_{\rm A}&=\Omega+\alpha\sin[\theta_{\rm A}(t)-\phi_{\rm A}(t)], 
\label{time_evolution_phase1}\\
\dot{\theta}_{\rm B}&=\Omega+\alpha\sin[\theta_{\rm B}(t)-\phi_{\rm B}(t)],
\label{time_evolution_phase2}
\end{align}
where $\Omega$ is the constant frequency, 
$\alpha$ is the coupling parameter describing the strength of synchronization, 
and $\phi_{\rm A}$ and $\phi_{\rm B}$ are the mechanical phases as explained below.

To define the above mechanical phases for a three-sphere model, it is convenient to introduce 
the following spring lengths $u_{\rm A}$ and $u_{\rm B}$ with respect to $\ell$:

\begin{align}
u_{\rm A}(t)& =x_2(t)-x_1(t)-\ell,
\label{uA} \\
u_{\rm B}(t)& =x_3(t)-x_2(t)-\ell.
\label{uAuB}
\end{align}
Obviously, these quantities are related to the sphere velocities as 
$\dot u_{\rm A} = \dot x_2- \dot x_1$ and $\dot u_{\rm B} = \dot x_3- \dot x_2$.
Then the time-dependent mechanical phases $\phi_{\rm A}$ and $\phi_{\rm B}$ are  
introduced by the relative positions and the velocities of the spheres as 

\begin{align}
\cos \phi_{\rm A}&=u_{\rm A}/D_{\rm A},~~ 
\sin \phi_{\rm A}=-\dot{u}_{\rm A}/(\Omega D_{\rm A}), 
\label{phiA} \\
\cos \phi_{\rm B}&=u_{\rm B}/D_{\rm B},~~
\sin \phi_{\rm B}=-\dot{u}_{\rm B}/(\Omega D_{\rm B}), 
\label{phiB}
\end{align}
where $D_{\rm A(B)}=[u_{\rm A(B)}^{2}+(\dot{u}_{\rm A(B)}/\Omega)^2]^{1/2}$. 
Physically, the mechanical phase $\phi_{\rm A(B)}$ specifies the position in the phase space
of a micromachine spanned by $u_{\rm A(B)}$ and $\dot{u}_{\rm A(B)}$ as shown in fig.~\ref{phasedynamics}.

\begin{figure}[tbh]
\centering
\includegraphics[scale=0.23]{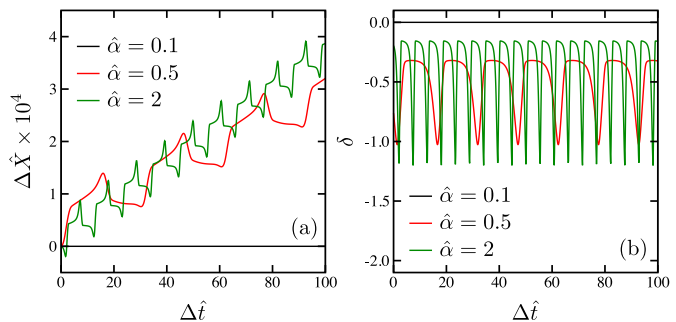}
\caption{
The plots of (a) dimensionless center-of-mass position difference 
$\Delta \hat{X}=\hat{X}(\hat{t}+\Delta\hat{t})-\hat{X}(\hat{t})$ and (b) the phase 
difference $\delta=\theta_{\rm B}-\theta_{\rm A}$ between the oscillations in the natural lengths 
as a function of dimensionless time difference $\Delta \hat{t}$ measured from 
$\hat{t}=199,900$ when $\hat{\Omega}=0.1$ and $\delta_0=-\pi/2$.
In both plots, the dimensionless coupling parameter is chosen as $\hat{\alpha}=0.1$ (black),  
$0.5$  (red), and $2$ (green).
The average steady state velocity $V_\infty$ is obtained by fitting with a straight line, 
whereas the steady state phase difference $\delta_\infty$ is obtained by averaging over a cycle 
in the oscillations of $\delta$.}
\label{fig:deltaXdeltat}
\end{figure}

The above equations complete our model for an autonomous three-sphere microswimmer.
In this letter, we shall consider the case of $\alpha \ge 0$.
Then the physical meaning of eqs.~(\ref{time_evolution_phase1}) and (\ref{time_evolution_phase2})
is that the phase $\theta_{\rm A}$ ($\theta_{\rm B}$) for the natural length and the mechanical phase 
$\phi_{\rm A}$ ($\phi_{\rm B}$) tend to be different due to the coupling term, as schematically shown 
in fig.~\ref{phasedynamics}.
Since the middle sphere is connected to the other two spheres, our model contains a feedback mechanism 
that regulates the dynamics of the two natural lengths $\ell_{\rm A}$ and $\ell_{\rm B}$.
Such a coupling effect in the spring motions gives rise to a non-reciprocal body motion and results in an 
autonomous locomotion of a microswimmer.
Although $\alpha$ in eqs.~(\ref{time_evolution_phase1}) and (\ref{time_evolution_phase2}) 
can be different, we shall first stick to the symmetric case for the sake of simplicity.
In general, the other quantities such as $K$, $\ell$, and $\Omega$ can also be asymmetric.

Let us define the time-dependent phase difference between the oscillations in the natural 
lengths by  
$\delta (t)=\theta_{\rm B}(t)-\theta_{\rm A}(t)$.
When $\alpha=0$, the present model reduces to that of the original elastic 
microswimmer~\cite{Yasuda17,Kuroda19}.
In this limit, we have $\theta_{\rm A}(t) = \Omega t$ and 
$\theta_{\rm B}(t) = \Omega t + \delta_0$, 
where $\delta_0=\delta(0)$ is the initial phase difference.
According to Purcell's scallop theorem~\cite{Purcell77,Lauga11}, an elastic microswimmer can 
exhibit a directed motion when $\delta_0 \neq 0, \pm \pi$, i.e., a nonreciprocal motion.
Hence the initial phase difference $\delta_0$ and the frequency $\Omega$ fully 
determines the average velocity of locomotion when $\alpha=0$~\cite{Yasuda17,Kuroda19}.

When the coupling effect is present, however, we show that a stable phase difference $\delta$
controls the dynamics of a micromachine irrespective of its initial value $\delta_0$.
Moreover, the transition to a nonreciprocal motion as well as the average velocity can be 
precisely tuned by the coupling parameter $\alpha$ and it is not solely fixed by the externally 
given frequency $\Omega$ as in the previous 
models~\cite{Golestanian04,Golestanian08,Yasuda17,Kuroda19}.

For numerical simulations, it is convenient to introduce a characteristic time scale defined by

\begin{equation}
\tau=\frac{6\pi\eta a}{K},
\label{calacteristictime}
\end{equation}
which represents the spring relaxation time.
Then we use $\ell$ to scale all the relevant lengths (such as $x_i$, $a$, and $d$) and employ $\tau$ to 
scale the quantities related to time (such as $\Omega$ and $\alpha$).
All the dimensionless variables and parameters are written with a hat such as $\hat{x}_i=x_i/\ell$, 
$\hat{\Omega}=\Omega \tau$, and $\hat{\alpha}=\alpha \tau$.

%%%%%%%%%%%%%%%
\section{Simulation results}
%%%%%%%%%%%%%%%

\begin{figure}[tbh]
\centering
\includegraphics[scale=0.23]{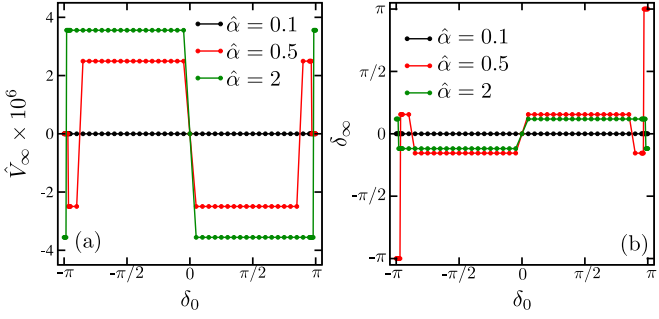}
\caption{
The plots of (a) dimensionless stationary velocity $\hat{V}_\infty$ and (b) stationary phase difference 
$\delta_\infty$ as a function of the initial phase difference $\delta_0$ ($-\pi \le \delta_0 \le \pi$)
when $\hat{\Omega}=0.1$.
In both plots, the dimensionless coupling parameter is chosen as $\hat{\alpha}=0.1$ (black),  
$0.5$  (red), and $2$ (green).
Notice that, for each color, there are multiple data points close to $\delta_0=\pm \pi$. 
}
\label{fig:V-delta}
\end{figure}

\begin{figure}[tbh]
\centering
\includegraphics[scale=0.23]{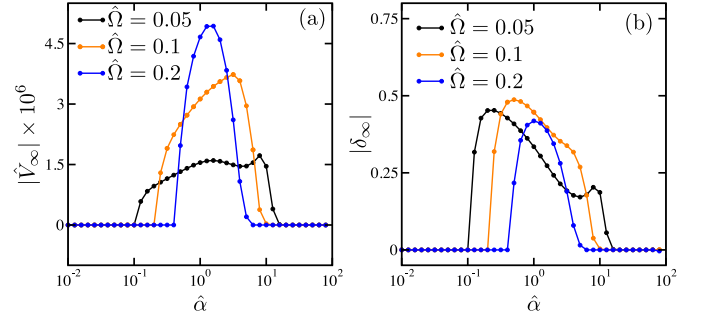}
\caption{
The plots of (a) dimensionless stationary velocity $\vert \hat{V}_\infty \vert$ and (b) stationary phase difference 
$\vert \delta_\infty \vert$ as a function of the dimensionless coupling parameter $\hat{\alpha}$.
In both plots, the dimensionless frequency is chosen as $\hat{\Omega}=0.05$ (black), $0.1$  (orange), 
and $0.2$ (blue), while $\delta_0=-\pi/2$ is fixed.
There is a lower critical value  $\alpha_{\rm c}$ above which both  $\vert V_\infty \vert$ and 
$\vert \delta_{\infty} \vert$ become nonzero. 
$\vert V_\infty \vert$ and $\vert \delta_\infty \vert$ take maximum values 
at $\alpha_{\rm m}>\alpha_{\rm c}$, and they vanish for large $\alpha$.
}
\label{fig:V-alpha}
\end{figure}

First we have performed computer simulations by numerically solving eq.~(\ref{ssokudo}) together 
with eqs.~(\ref{ellA})--(\ref{phiB}) with the use of the Euler's method. 
The parameters to characterize the swimmer size are chosen as  $\hat{a}=0.01$ and $\hat{d}=0.1$,
satisfying the conditions $a, d \ll \ell$.
Concerning the initial conditions, we put the three spheres at $\hat{x}_1(0)=-1$, $\hat{x}_2(0)=0$, 
and $\hat{x}_3(0)=1$, whereas the initial phase difference, $\delta_0$, is varied within 
the range $-\pi \leq \delta_0 \leq \pi$.
In the present work, we focus on the low-frequency regime, $\hat{\Omega}<1$.
The following simulation results do not depend on the initial positions of the three spheres.

\begin{figure}[tbh]
\centering
\includegraphics[scale=0.3]{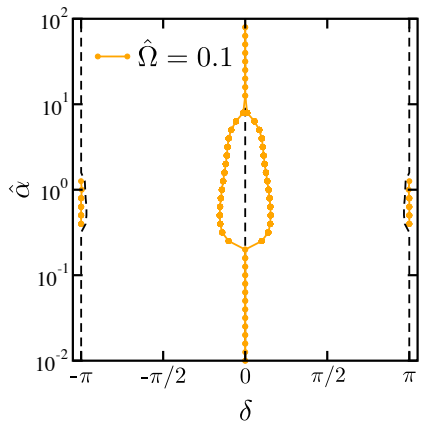}
\caption{The numerically obtained stability diagram in the plane of the phase difference $\delta$ and 
the dimensionless coupling parameter $\hat{\alpha}$ when $\hat{\Omega}=0.1$.
The orange circles indicate stable fixed points where $\dot{\delta}=0$ holds. 
For $\hat{\alpha} > \hat{\alpha}_{\rm c} \approx 0.2$, the stable point $\delta=0$ bifurcates into 
two stable fixed points.
These nonzero fixed points correspond to nonreciprocal motions 
($\delta \neq 0, \pm \pi$), leading to a nonzero velocity $V_\infty$.   
There are other stable fixed points at $\delta = \pm \pi$ when $0.39 \le \hat{\alpha} \le 1.58$.
For larger $\hat{\alpha}$, $\delta =0$ becomes stable again for $\hat{\alpha}\ge 12.5$.
The dashed lines indicate the numerically determined separatrices which are obtained 
by comparing the initial phase difference $\delta_{0}$ and the stationary phase difference 
$\delta_{\infty}$.
}
\label{fig:sim-phasediagram}
\end{figure}

In fig.~\ref{fig:d-time}, we plot the dimensionless center-of-mass position $\hat{X}$ as a function of 
time $\hat{t}$ for different values of the coupling parameter $\hat{\alpha}$ when (a) $\delta_0=-\pi/2$ 
and (b) $\delta_0=-39\pi/40$, whereas the frequency is fixed to $\hat{\Omega}=0.1$.
Although $\hat{X}$ also oscillates in time at much smaller time scales, as shown later in 
fig.~\ref{fig:deltaXdeltat}(a), one can extract an average steady state velocity $V_\infty$ in the long-time 
limit by fitting with a straight line.
We estimate such an average velocity $V_\infty$ for each curve in fig.~\ref{fig:d-time}, and regard it 
as an autonomously determined steady state velocity.
For $\hat{\alpha}=0.1$ (black),  $V_\infty$ vanishes both in figs.~\ref{fig:d-time}(a) and (b).
For $\hat{\alpha}=0.5$ (red), on the other hand, $V_\infty$ is finite in fig.~\ref{fig:d-time}(a) 
but vanishes in (b). 
In this case, the steady state velocity depends on $\delta_0$. 
For $\hat{\alpha}=2$ (green), $V_\infty$ is the same between figs.~\ref{fig:d-time}(a) and (b), 
showing that $V_\infty$ does not depend on the initial phase difference $\delta_0$ 
although the sign of $V_\infty$ can change as we show later in fig.~\ref{fig:V-delta}(a).

In figs.~\ref{fig:deltaXdeltat}(a) and (b), the behaviors of the center-of-mass position difference 
$\Delta \hat{X}=\hat{X}(\hat{t}+\Delta\hat{t})-\hat{X}(\hat{t})$ and the phase difference 
$\delta$ at much smaller time scales are plotted, respectively, as a function of the 
time difference $0 \le \Delta \hat{t} \le 100$ measured after $\hat{t}=199,900$.
Since the other parameters are $\delta_0=-\pi/2$ and $\hat{\Omega}=0.1$, fig.~\ref{fig:deltaXdeltat}(a) 
is the magnification of fig.~\ref{fig:d-time}(a) in the long-time limit after the steady state has 
been reached.
It is important to note that both $\Delta \hat{X}$ and $\delta$ exhibit oscillatory behaviors whose
period becomes smaller as $\alpha$ is increased. 
Such a change in the period is consistent with a perturbation expansion 
of eqs.~(\ref{time_evolution_phase1}) and (\ref{time_evolution_phase2}) in terms of $\alpha$, as we 
shall explain later.
In fig.~\ref{fig:deltaXdeltat}(b), the phase difference $\delta$ oscillates around a constant value that 
can be regarded as the steady state phase difference $\delta_\infty$.
Here, we define $\delta_\infty$ as the average over a cycle in the oscillations of $\delta$.
Since there is always a well-defined steady state for a given set of parameters,
further simulations have been performed for different values of $\delta_0$ to  
investigate the behaviors of $V_\infty$ and $\delta_\infty$ systematically.

Fixing the frequency to $\hat{\Omega}=0.1$, we plot in figs.~\ref{fig:V-delta}(a) and (b)
the steady state velocity $\hat{V}_\infty$ and the phase difference $\delta_\infty$,
respectively, as a function of the initial phase difference $\delta_0$ for the range $-\pi \le \delta_0 \le \pi$.
Different colors indicate different $\hat{\alpha}$ values as we have used in fig.~\ref{fig:d-time}.
In fig.~\ref{fig:V-delta}(a), we see that $V_\infty$ either vanishes or takes a nonzero constant value 
within a certain range of $\delta_0$.
This means that, under certain conditions, the proposed microswimmer can autonomously determine its 
steady state velocity as well as the phase difference.
We also see that $\hat{V}_\infty$ changes its sign at $\delta_0=0$ although the absolute value
is the same.
The sign of $\hat{V}_\infty$ and $\delta_\infty$ also changes for $\hat{\alpha}=0.5$ (red) 
and $2$ (green) when $\delta_0$ becomes close to $\pm \pi$.

For $\hat{\alpha}=0.5$ (red) in fig.~\ref{fig:V-delta}(a), the velocity $V_\infty$ tends to vanish when 
the initial phase difference $\delta_0$ is close to $\pm \pi$. 
In this situation, we see in fig.~\ref{fig:V-delta}(b) that the steady state phase difference approaches 
$\delta_\infty = \pm \pi$, i.e., a reciprocal motion.
When $V_\infty$ is finite in fig.~\ref{fig:V-delta}(a) for $\hat{\alpha}=0.5$ (red) and $2$ (green), 
on the other hand, the corresponding phase difference is $\delta_\infty \neq 0, \pm \pi$, i.e., 
a nonreciprocal motion. 
These results are in accordance with Purcell's scallop theorem~\cite{Purcell77,Lauga11}.
A more detailed discussion concerning the stability of the phase difference will be given 
later in fig.~\ref{fig:sim-phasediagram}.
When $\hat{\alpha}=0.5$, there are two stable fixed points; one with finite $V_\infty$ and the 
other with vanishing $V_\infty$.

In figs.~\ref{fig:V-alpha}(a) and (b), we plot $\vert \hat{V}_\infty \vert$ and 
$\vert \delta_\infty \vert$, respectively, as a function of $\hat{\alpha}$ for different 
frequencies $\hat{\Omega}=0.05$, $0.1$, and $0.2$.
To make these plots, we have used $\delta_0=-\pi/2$.
When $\hat{\Omega}=0.1$ (orange), for example, there is a finite critical value of $\hat{\alpha}_{\rm c} 
\approx 0.2 $ above which $\vert \hat{V}_\infty \vert$ and $\vert \delta_\infty \vert$ become nonzero.
For $\hat{\alpha} < \hat{\alpha}_{\rm c}$, on the other hand, both $\vert \hat{V}_\infty \vert$ and 
$\vert \delta_\infty \vert$ vanish. 
The existence of such a finite critical value $\hat{\alpha}_{\rm c}$ is a nontrivial outcome of 
the present model.
When $\hat{\alpha}$ is very large, such as $\hat{\alpha}\ge 12.5$ for $\hat{\Omega}=0.1$,
both $\vert \hat{V}_\infty \vert$ and $\vert \delta_\infty \vert$ vanish again. 
Hence autonomous locomotion can be achieved for a finite range of the coupling parameter 
$\hat{\alpha}$.
Such a behavior is common for other frequencies $\hat{\Omega}$.

Moreover, it is interesting to note that $\vert V_\infty \vert$ takes a  
maximum values such as at $\hat{\alpha}_{\rm m} \approx 5.0$ when $\hat{\Omega}=0.1$.
Hence the present autonomous microswimmer can maximize its velocity by tuning the coupling 
parameter $\alpha$.
Notice that both $\hat{\alpha}_{\rm c}$ and $\hat{\alpha}_{\rm m}$ depend on the frequency 
$\hat{\Omega}$, and they are not universal quantities.
However, it is worth mentioning that we find the relation $\hat{\alpha}_{\rm c}/\hat{\Omega} \approx 2$
for all $\hat{\Omega}$ chosen in our simulations.

From these simulation results, one can discuss the stability of the relative phase difference $\delta$. 
Comparing its initial value $\delta_0$ and the steady state value $\delta_\infty$, we can identify the 
stable fixed points.
Such a stability diagram for $\hat{\Omega}=0.1$ is presented in fig.~\ref{fig:sim-phasediagram} in 
the plane of $\delta$ and $\hat{\alpha}$, describing the bifurcation structure of the present model.
The orange circles indicate the stable fixed points where $\dot{\delta}=0$ holds.
In fig.~\ref{fig:sim-phasediagram} we have also plotted the numerically determined 
separatrices by the dashed lines. 
These points are obtained by comparing the initial phase difference $\delta_{0}$ and the stationary 
phase difference $\delta_{\infty}$.
Hence they are not mathematically obtained rigorous unstable points.

For $\hat{\alpha} < \hat{\alpha}_{\rm c} \approx 0.2$, the stable fixed points exist
only at $\delta =0$.
As $\hat{\alpha}$ is increased, the stable point at $\delta =0$ bifurcates into two stable fixed 
points for $\hat{\alpha} > \hat{\alpha}_{\rm c}$, whereas $\delta =0$ becomes unstable.
The stable fixed points at nonzero $\delta$ correspond to nonreciprocal motions, 
leading to finite $V_\infty$ and $\delta_\infty$.
When $\hat{\alpha}$ satisfies $0.39 \le \hat{\alpha} \le 1.58$, the two stable fixed points at 
$\delta = \pm \pi$ appear.
These new fixed points result in a reciprocal motion, prohibiting the locomotion of a microswimmer.  
For larger coupling parameter $\hat{\alpha}\ge 12.5$, $\delta =0$ becomes stable again.
Although such a stability diagram depends on $\hat{\Omega}$, the general 
structure of the bifurcation diagram remains the same.

%%%%%%%%%%%%%%%%%%%%%%%%%%%%%%%
\section{Linear stability analysis in the weak coupling limit}
%%%%%%%%%%%%%%%%%%%%%%%%%%%%%%%

Although the equations of motion of our model are highly nonlinear, we can analytically investigate 
the linear stability of the phase difference $\delta$ when $\hat{\alpha}$ is small enough.
In other words, we consider the case $\hat{\alpha} < \hat{\alpha}_{\rm c}$ in fig.~\ref{fig:sim-phasediagram},
for which $\delta=0$ is the only stable point and a microswimmer exhibits a reciprocal motion.
Here, we shall express $\dot{\delta}$ in terms of $\delta$ and perform a stability analysis.
Starting from eq.~(\ref{ssokudo}), we first neglect hydrodynamic interactions by considering the
case $a \ll \ell$.  
Then the equations of motion for  $u_{\rm A}$ and $u_{\rm B}$ [see eqs.~(\ref{uA}) and (\ref{uAuB})] 
are approximated as 

\begin{align}
\dot{u}_{\rm A}& \approx \frac{1}{\tau}
\left[ -2(u_{\rm A}-d\cos\theta_{\rm A})
+(u_{\rm B}-d\cos\theta_{\rm B}) \right],
\label{eqofuadot} \\
\dot{u}_{\rm B}& \approx \frac{1}{\tau}
\left[(u_{\rm A}-d\cos\theta_{\rm A})
-2(u_{\rm B}-d\cos\theta_{\rm B})\right],
\label{eqofubdot}
\end{align}
where $\tau$ is the spring relaxation time introduced in eq.~(\ref{calacteristictime}).

When the coupling parameter $\alpha$ is small enough in eqs.~(\ref{time_evolution_phase1}) 
and (\ref{time_evolution_phase2}), one can assume that $\delta$ is almost constant and
the phases of the two natural lengths can be approximated as $\theta_{\rm A}(t) \approx \Omega t$ 
and $\theta_{\rm B}(t) \approx \Omega t +\delta$.
According to ref.~\cite{Yasuda17}, the coupled linear equations in eqs.~(\ref{eqofuadot}) and 
(\ref{eqofubdot}) can  be solved in the frequency domain.
By performing the inverse Fourier transform, we have 

\begin{align}
& u_{\rm A}(t)\approx\frac{d}{9+10\hat{\Omega}^2+\hat{\Omega}^4}
\nonumber \\ 
& \times \left[
(9+5\hat{\Omega}^2)\cos(\Omega t)-4\hat{\Omega}^2\cos(\Omega t+\delta) 
\right. \nonumber \\
&\left. +(6\hat{\Omega}+2\hat{\Omega}^3)\sin(\Omega t)
+(3\hat{\Omega}-\hat{\Omega}^3)\sin(\Omega t+\delta)
\right],
\label{ua_solution} \\
&u_{\rm B}(t)\approx \frac{d}{9+10\hat{\Omega}^2+\hat{\Omega}^4}
\nonumber \\
& \times \left[
-4\hat{\Omega}^2\cos(\Omega t)+(9+5\hat{\Omega}^2)\cos(\Omega t+\delta) 
\right. \nonumber \\
&\left. +(3\hat{\Omega}-\hat{\Omega}^3)\sin(\Omega t)
+(6\hat{\Omega}+2\hat{\Omega}^3)\sin(\Omega t+\delta) 
\right],
\label{ub_solution}
\end{align}
where $\hat{\Omega}=\Omega \tau$ as before and $\delta$ here is a constant.

\begin{figure}[tbh]
\centering
\includegraphics[scale=0.3]{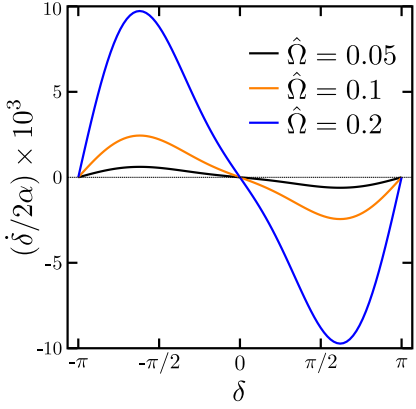}
\caption{
The plot of $\dot{\delta}/(2\alpha)$ [see eq.~(\ref{analyticalsolution})] as a function of $\delta$ for 
different dimensionless frequencies $\hat{\Omega}=0.05$ (black), $0.1$  (orange), and $0.2$ (blue).
Here $\delta=0$ corresponds to the stable fixed point, while $\delta=\pm \pi$ are unstable ones.
This result is in agreement with the stability diagram in fig.~\ref{fig:sim-phasediagram} when  
$\alpha < \alpha_{\rm c}$. 
}
\label{fig:analyticalsolution}
\end{figure}

The above results can be inserted into eqs.~(\ref{phiA}) and (\ref{phiB}) to obtain $\phi_{\rm A}(t)$ 
and $\phi_{\rm B}(t)$, respectively.
Considering the low-frequency limit, $\hat{\Omega} \ll 1$, we expand $\phi_{\rm A}$ and $\phi_{\rm B}$ 
up to the second order in $\hat{\Omega}$:

\begin{align}
\phi_{\rm A}(t)&\approx \Omega t -\frac{2+\cos \delta}{3}\hat{\Omega}
-\frac{(2-\cos \delta)\sin \delta}{9} \hat{\Omega}^2,
\label{solved_phi12} \\
\phi_{\rm B}(t)&\approx \Omega t +\delta -\frac{2+\cos \delta}{3}\hat{\Omega}
+\frac{(2-\cos \delta)\sin \delta}{9} \hat{\Omega}^2.
\label{solved_phi23}
\end{align}
Then we substitute these expressions into eqs.~(\ref{time_evolution_phase1}) 
and (\ref{time_evolution_phase2}) to obtain
\begin{align}
\dot{\theta}_{\mathrm A} &\approx \Omega + \alpha \sin \left [ \frac{2+\cos \delta}{3}\hat{\Omega} + \frac{(2-\cos \delta)\sin \delta}{9} \hat{\Omega}^2 \right ],
\label{modified_time_evolution_phase1} \\
\dot{\theta}_{\mathrm B} &\approx \Omega + \alpha \sin \left [ \frac{2+\cos \delta}{3}\hat{\Omega} - \frac{(2-\cos \delta)\sin \delta}{9} \hat{\Omega}^2 \right ],
\label{modified_time_evolution_phase2}
\end{align}
where we have used the approximations $\theta_{\rm A}(t) \approx \Omega t$ and 
$\theta_{\rm B}(t) \approx \Omega t +\delta$.
The stability of $\delta$ can be discussed in terms of $\dot{\delta}=\dot{\theta}_{\rm B} - \dot{\theta}_{\rm A}$ 
that is given by 
\begin{align}
\frac{\dot{\delta}}{2\alpha} 
\approx \sin\left[ \frac{(-2+\cos \delta)\sin \delta}{9} \hat{\Omega}^{2}\right]
\cos\left[\frac{2+\cos \delta}{3} \hat{\Omega}\right].
\label{analyticalsolution}
\end{align}

In fig.~\ref{fig:analyticalsolution}, we plot $\dot{\delta}/(2\alpha)$ as a function of $\delta$ for 
$\hat{\Omega}=0.05$, $0.1$, and $0.2$.
We first note that eq.~(\ref{analyticalsolution}) is an odd function of $\delta$. 
Since $\dot{\delta}>0$ for $\delta<0$ and $\dot{\delta}<0$ for $\delta>0$, 
we find that $\delta=0$ is a stable fixed point. 
This result is in accordance with the stability diagram in fig.~\ref{fig:sim-phasediagram} when 
$\alpha < \alpha_{\rm c}$. 
The slope at $\delta=0$ becomes steeper as $\hat{\Omega}$ is increased.  
We also see that $\delta=\pm \pi$ are the unstable fixed points.

Up to the first order in $\hat{\Omega}$, we see in eqs.~(\ref{solved_phi12}) and (\ref{solved_phi23}) 
that the mechanical phases $\phi_\mathrm{A}$ and $\phi_\mathrm{B}$ are delayed with respect to those 
of the natural length $\theta_\mathrm{A}$ and $\theta_\mathrm{B}$, respectively.
Thus, the time evolutions of $\theta_\mathrm{A}$ and $\theta_\mathrm{B}$ are accelerated by the second 
terms in eqs.~(\ref{modified_time_evolution_phase1}) and (\ref{modified_time_evolution_phase2}) 
that are controlled by $\alpha$.
Hence, as $\alpha$ is increased, the oscillation frequencies of $\Delta \hat{X}$ and $\delta$ become 
larger than the original spring frequency $\Omega$.
Such a change of the oscillation frequency was shown in fig.~\ref{fig:deltaXdeltat}(b).

Within the present approximation, however, we cannot analytically predict the critical value 
$\alpha_{\rm c}$ nor the stable fixed points at nonzero $\delta$ for $\alpha > \alpha_{\rm c}$.
The difficulty arises because eqs.~(\ref{ua_solution}) and (\ref{ub_solution}) are correct 
only for small $\alpha$.
Moreover, hydrodynamic interactions, which are neglected in the above analysis, need to be further
taken into account to fully discuss the bifurcation structure of the model.

It is worth mentioning, however, that the stability of the phase difference $\delta$ 
can be determined even in the absence of hydrodynamic interactions as we have discussed 
in this section. 
For a three-sphere model of \textit{Chlamydomonas}, it was shown that hydrodynamic interactions
contribute little to synchronization~\cite{Friedrich12}.
In the present model as well as in the previous models~\cite{Golestanian04,Golestanian08,Yasuda17,Kuroda19}, 
hydrodynamic interactions play an essential role for the locomotion of a three-sphere 
microswimmer.

%%%%%%%%%%%%%%%%%%
\section{Summary and discussion}
%%%%%%%%%%%%%%%%%%

In this letter, we have proposed a model of an autonomous three-sphere microswimmer by considering 
a coupling effect between the two natural lengths of an elastic microswimmer~\cite{Yasuda17}.   
Our model is motivated by the previous models for synchronization phenomena in coupled 
oscillator systems~\cite{Kuramoto84,Pikovsky01,Strogatz03,Ritort05}.
Performing numerical simulations, we have shown that a microswimmer can acquire a nonzero
steady state velocity $V_\infty$ that is almost independent of the initial phase difference $\delta_0$
[see fig.~\ref{fig:V-delta}(a)]. 
The corresponding phase difference $\delta_\infty$ between the oscillations in the natural lengths
becomes also finite [see fig.~\ref{fig:V-delta}(b)], which is consistent with Purcell's scallop theorem for 
microswimmers in a viscous fluid~\cite{Purcell77,Lauga11}.

We have explored in detail the dependencies of  $V_\infty$ and $\delta_\infty$ on the coupling 
parameter $\alpha$ and the frequency $\Omega$.
We find that both $\vert V_\infty \vert$ and $\vert \delta_\infty \vert$ take nonzero values 
for $\alpha > \alpha_{\rm c}$, and they also show maximum values at 
$\alpha_{\rm m}$ (fig.~\ref{fig:V-alpha}).
There is a finite range of $\alpha$ for which a microswimmer can have an autonomous directed motion.
We have also analyzed the stability of the phase difference $\delta$ by constructing a stability 
diagram (fig.~\ref{fig:sim-phasediagram}).
This result has been analytically confirmed in the limit of small $\alpha$ (fig.~\ref{fig:analyticalsolution}).

In the present work, we have discussed the case when the frequency $\Omega$ is small 
enough, i.e, $\hat{\Omega} <1$. 
When $\Omega$ is made larger, the difference between $\theta_{\rm A(B)}$ (phases of the 
natural lengths) and $\phi_{\rm A(B)}$ (mechanical phases) becomes also larger.
In the original elastic microswimmer without any coupling effect, it was shown that the average 
velocity decreases with increasing frequency in the high-frequency limit due to the intrinsic spring 
relaxation dynamics~\cite{Yasuda17,Kuroda19}.
Such a reduction of the velocity in the high-frequency regime also occurs for a three-sphere 
microswimmer moving in a viscoelastic medium~\cite{Yasuda17-2,Yasuda18,Yasuda20}.
Our future numerical and analytical studies include not only the high-frequency behavior of the 
model but also the case of $\alpha < 0$.

%%%%%%%%%%
\acknowledgments
%%%%%%%%%%

We thank T.\ Kato for useful discussions.
Y.K.\ acknowledges support by a Grant-in-Aid for JSPS Fellows (Grant No.\ JP19J00365) from the Japan Society 
for the Promotion of Science (JSPS). 
Y.H.\ acknowledges support by a Grant-in-Aid for JSPS Fellows (Grant No.\ 19J20271) from the JSPS.
K.Y.\ acknowledges support by a Grant-in-Aid for JSPS Fellows (Grant No.\ 18J21231) from the JSPS.
Y.K., H.K.\ and S.K.\ acknowledge support by a Grant-in-Aid for Scientific Research (C) (Grant No.\ 19K03765) 
from the JSPS.
S.K.\ further acknowledges support by a Grant-in-Aid for Scientific Research (C) (Grant No.\ 18K03567) 
from the JSPS, and support by a Grant-in-Aid for Scientific Research on Innovative Areas
``Information Physics of Living Matters'' (Grant No.\ 20H05538) from the Ministry of Education, Culture, 
Sports, Science and Technology of Japan.

%%%%%%%%%%%%%%%


\begin{thebibliography}{0}
%%%%%%%%%%%%%%%

\bibitem{Lauga09}
\Name{Lauga E. \and Powers T. R.}
\REVIEW{Rep. Prog. Phys.}{72}{2009}{096601}.

\bibitem{Purcell77}
\Name{Purcell E. M.} 
\REVIEW{Am. J. Phys.}{45}{1977}{3}. 

\bibitem{Lauga11}
\Name{Lauga E.} 
\REVIEW{Soft Matter}{7}{2011}{3060}. 

\bibitem{Golestanian04}
\Name{Najafi A. \and Golestanian R.}
\REVIEW{Phys. Rev. E}{69}{2004}{062901}.

\bibitem{Golestanian08}
\Name{Golestanian R. \and Ajdari A.}
\REVIEW{Phys. Rev. E}{77}{2008}{036308}.

\bibitem{Leoni09}
\Name{Leoni M., Kotar J., Bassetti B., Cicuta P. \and Lagomarsino M. C.}
\REVIEW{Soft Matter}{5}{2009}{472}.

\bibitem{Grosjean16}
\Name{Grosjean G., Hubert M., Lagubeau G. \and Vandewalle N.}
\REVIEW{Phys. Rev. E}{94}{2016}{021101(R)}.

\bibitem{Grosjean18}
\Name{Grosjean G., Hubert M. \and Vandewalle N.}
\REVIEW{Adv. Colloid Interface Sci.}{255}{2018}{84}.

\bibitem{Yasuda17}
\Name{Yasuda K., Hosaka Y., Kuroda M., Okamoto R. \and Komura S.}
\REVIEW{J. Phys. Soc. Jpn.}{86}{2017}{093801}.

\bibitem{Dunkel09}
\Name{Dunkel J. \and Zaid I. M.}
\REVIEW{Phys. Rev. E}{80}{2009}{021903}.

\bibitem{Pande15}
\Name{Pande J. \and Smith A.-S.}
\REVIEW{Soft Matter}{11}{2015}{2364}.

\bibitem{Pande17}
\Name{Pande J., Merchant L., Kr\"{u}ger T., Harting J. \and Smith A.-S.}
\REVIEW{New J. Phys.}{19}{2017}{053024}.

\bibitem{Kuroda19}
\Name{Kuroda M., Yasuda K. \and Komura S.}
\REVIEW{J. Phys. Soc. Jpn.}{88}{2019}{054804}.

\bibitem{Hosaka17}
\Name{Hosaka Y., Yasuda K., Sou I., Okamoto R. \and Komura S.}
\REVIEW{J. Phys. Soc. Jpn.}{86}{2017}{113801}.

\bibitem{Sou19}
\Name{Sou I., Hosaka Y., Yasuda K. \and Komura S.}
\REVIEW{Phys. Rev. E}{100}{2019}{022607}.

\bibitem{Sou20}
\Name{Sou I., Hosaka Y., Yasuda K. \and Komura S.}
\REVIEW{Physica A}{562}{2021}{125277}.

\bibitem{Owaki12}
\Name{Owaki D., Kano T., Nagasawa K., Tero A. \and Ishiguro A.}
\REVIEW{J. R. Soc. Interface}{10}{2012}{20120669}.

\bibitem{Fukuhara18}
\Name{Fukuhara A., Owaki D., Kano T., Kobayashi R. \and Ishiguro A.}
\REVIEW{Adv. Robot.}{32}{2018}{794}.

\bibitem{Kuramoto84}
\Name{Kuramoto Y.}
\Book{Chemical Oscillations, Waves, and Turbulence}
\Publ{Springer-Verlag, New York} \Year{1984}.

\bibitem{Pikovsky01}
\Name{Pikovsky A., Rosenblum M. \and Kurths J.}
\Book{Synchronization: A Universal Concept in Nonlinear Sciences}
\Publ{Cambridge University Press, Cambridge} \Year{2001}.

\bibitem{Strogatz03}
\Name{Strogatz S. H.}
\Book{Sync: The Emerging Science of Spontaneous Order}
\Publ{Hyperion, New York} \Year{2003}.

\bibitem{Ritort05}
\Name{Acebr\'{o}n J. A., Bonilla L. L., P\'{e}rez Vicente C. J., Ritort F. \and Spigler R.}
\REVIEW{Rev. Mod. Phys.}{77}{2005}{137}.

\bibitem{Golestanian11}
\Name{Golestanian R., Yeomans J. M. \and Uchida N.}
\REVIEW{Soft Matter}{7}{2011}{3074}.

\bibitem{Uchida17}
\Name{Uchida N., Golestanian R. \and Bennett R. R.}
\REVIEW{J. Phys. Soc. Jpn.}{86}{2017}{101007}.

\bibitem{Polin09}
\Name{Polin M., Tuval I., Drescher K., Gollub J. P. \and Goldstein R. E.}
\REVIEW{Science}{325}{2009}{487}.

\bibitem{Goldstein09}
\Name{Goldstein R. E., Polin M. \and Tuval I.}
\REVIEW{Phys. Rev. Lett.}{103}{2009}{168103}.

\bibitem{Friedrich12}
\Name{Friedrich B. M. \and J\"{u}licher}
\REVIEW{Phys. Rev. Lett.}{109}{2012}{138102}.

\bibitem{Bennett13}
\Name{Bennett R. R. \and Golestanian R.}
\REVIEW{Phys. Rev. Lett.}{110}{2013}{148102}.


\bibitem{Yasuda17-2}
\Name{Yasuda K., Okamoto R. \and Komura S.}
\REVIEW{J. Phys. Soc. Jpn.}{86}{2017}{043801}.

\bibitem{Yasuda18}
\Name{Yasuda K., Okamoto R. \and Komura S.}
\REVIEW{EPL}{123}{2018}{34002}.

\bibitem{Yasuda20}
\Name{Yasuda K., Kuroda M. \and Komura S.}
\REVIEW{Phys. Fluids}{32}{2020}{093102}.

\end{thebibliography}
\end{document}